**High Efficiency Thin Film Superlattice Thermoelectric Cooler Modules Enabled by Low Resistivity Contacts**

*Yuping He, François Léonard, Douglas L. Medlin, Nicholas Baldasaro, Dorota S. Temple[*], Philip Barletta[*], and Catalin D. Spataru[*]*

Dr. Y. He, Dr. F. Léonard, Dr. D.L. Medlin, Dr. C.D. Spataru
Sandia National Laboratories
Livermore, California, 94551 (USA)
E-mail: cdspata@sandia.gov
Mr. N. Baldasaro, Dr. D. S. Temple
RTI International
Research Triangle Park, North Carolina, 27709 (USA)
E-mail: temple@rti.org
Dr. P. Barletta
Micross Components
3021 East Cornwallis Road
Research Triangle Park, North Carolina, 27709 (USA)
E-mail: Philip.Barletta@micross.com

Abstract

V-telluride superlattice thin films have shown promising performance for on-chip cooling devices. Recent experimental studies have indicated that device performance is limited by the metal/semiconductor electrical contacts. One challenge in realizing a low resistivity contacts is the absence of fundamental knowledge of the physical and chemical properties of interfaces between metal and V-telluride materials. Here we present a combination of experimental and theoretical efforts to understand, design and harness low resistivity contacts to V-tellurides. *Ab initio* calculations are used to explore the effects of interfacial structure and chemical compositions on the electrical contacts, and an *ab initio* based macroscopic model is employed to predict the fundamental limit of contact resistivity as a function of both carrier concentration and temperature.

Under the guidance of theoretical studies, we develop an experimental approach to fabricate low resistivity metal contacts to V-telluride thin film superlattices, achieving a 100-fold reduction compared to previous work. Interfacial characterization and analysis using both scanning transmission electron microscopy and energy-dispersive x-ray spectroscopy show the unusual interfacial morphology and the potential for further improvement in contact resistivity. Finally, we harness the improved contacts to realize an improved high-performance thermoelectric cooling module.

V-telluride materials such as $Bi_2Te_3$ and $Sb_2Te_3$ have been of interest for thermoelectric applications[1-5] and recently for their topological insulator properties[6-9]. In particular, thin films of such materials have shown excellent performance for on-chip cooling[10-13] with even better performance predicted theoretically[14]. To realize the full promise of these materials, it is necessary to eliminate parasitics that can limit the performance. As such, metal contacts play an important role for thin-film thermoelectric (TE) devices, especially in high heat-flux applications (e.g. chip cooling) where low contact resistivity ($\rho_C$) is critical to device performance[15-17]. Recent work[18] has demonstrated low electrical contact resistivity $\rho_C$ in the range $1-2\times10^{-6}$ $\Omega cm^2$ in TE modules based on $(Bi,Sb)_2Te_3$ superlattices using the evaporation of Cr/Ni/Au to fabricate metal electrodes. For thin-film thermoelectric modules with the TE thickness < 2 μm, further reduction of $\rho_C$ to $10^{-8}$ $\Omega cm^2$ is needed for the contact resistivity to be a small fraction of the resistivity of the thermoelectric element itself[18].

Creating such low contact resistivities is challenging from a fabrication perspective[19], but also because little is known about the fundamental properties of metal contacts to these materials. For example, even basic properties such as the atomic and electronic structure of the metal/TE interface are largely unknown. This makes it difficult to optimize the contact resistivity and to establish the fundamental limits[20] that are possible. To address this challenge, we present an integrated theoretical and experimental effort towards understanding the limits of low-$\rho_C$ in realistic metal contacts to advanced TE materials. We present a new multiscale theoretical approach combining *ab initio* calculations and continuum mesoscopic models to investigate the structural, electronic and transport properties of electrical contacts to novel TE materials used in thin-film, superlattice V-telluride devices. We show that the nature of these semiconductor materials leads to unusual contact properties, such as strong n-type doping near the interface and interfacial atomic dipoles that completely determine the band-bending. We predict that significant improvement over previously reported experimental data is possible, and we present new experimental data that demonstrate a 100-fold reduction in contact resistivity. Detailed atomistic spatially-resolved measurements of the new contacts show that additional improvement should be possible. Importantly, we demonstrate that the reduction in contact resistivity can be harnessed to improve the thermoelectric efficiency of cooling modules.

To understand the electrical properties of contacts to $(Bi,Sb)_2Te_3$ materials and their realistic low $\rho_C$ limit we carried out a series of large scale *ab initio* calculations of the $Sb_2Te_3$-Cr interface illustrated in Fig. 1a. Cr was chosen because it has shown good adhesion properties to V-telluride materials[21, 22], while $Sb_2Te_3$ possesses lower electrical resistivity than $Bi_2Te_3$, possibly implying a lower contact resistivity. In addition, we also chose to simulate the p-type doped TE material since experimentally[18] it has higher contact resistivity compared to the n-type material, and thus greater gains might be expected from improvement of that contact. While metal inter-diffusion at metal/TE interfaces is known to arise[23] and Cr is a common dopant[24, 25] for $Sb_2Te_3$, such effects are beyond the scope of this work, and we therefore focus on the clean interface in our simulations.

We use *ab initio* calculations to simulate the details of the contacts. *Ab initio* calculations are necessary because the band alignment at metal/semiconductor interfaces is determined by the atomistic details of the interface. For example, it is well known that Fermi level pinning due to interface states leads to Schottky barriers even when the band alignment should give an Ohmic contact. The importance of *ab initio* calculations in determining contact resistivity has been demonstrated previously in the case of conventional semiconductors[26-29]. We perform the calculations using the VASP code[30, 31] to simulate realistic low-strained metal/TE interfaces using large supercells (thousands of atoms), which allows us to incorporate anti-site defects that yield p-type free carrier concentrations similar to those expected to maximize cooling performance. (See Supporting Information for additional details of the computation.) Figure 1b shows the calculated electronic structure of bulk undoped $Sb_2Te_3$. We find a quasi-direct band gap of 100 meV with both the VBM and CBM having six-fold valley degeneracy. From the calculated bandstructure we also calculated the density of modes[32], from which hole and electrons transport effective masses of ~ 0.25 $m_0$, where $m_0$ is the free electron mass, were obtained perpendicular to the Sb or Te layers. Using a dense k-point grid we have also calculated the carrier concentration (holes and electrons) as a function of position of Fermi level, as shown in Fig. 1c.

To simulate the p-doped semiconductor we created two anti-site defects (Te replaced by Sb) in our supercell. At this density of anti-site defects the Fermi level is located in the valence band, with no localized defect states in the bandgap. Given that Te has one more valence electron than Sb, each anti-site defect leads to one free hole in the degenerate doping regime. We thus realize a total doping concentration expected to lead to optimal TE properties, namely p=6.5 $\times 10^{19}$

holes/cm$^3$. Based on the carrier concentration results of Fig. 1c, the Fermi level lies 75 meV below the VBM for this doping level, as indicated with the dashed line in Fig. 1b.

The simplest model for the band alignment at the metal/semiconductor interface is the direct alignment due to the metal work function $\phi$ and the semiconductor electron affinity $\chi$. We thus calculated these two quantities from *ab initio* (see Supplementary Information for details), finding $\phi = 4.4$ eV and $\chi = 4.7$ eV as shown in Fig. 1a. This large $\chi$ implies that a number of metals will naturally prefer to form n-type contacts to Sb$_2$Te$_3$; in addition, the large electron affinity will lead to electron transfer from the metal to the semiconductor, which will dope the semiconductor n-type near the interface. This has important consequences for contacts to p-type materials since we can expect a p-n junction to form near the contact which will increase the contact resistivity.

Of course, the local atomic interactions at the interface can change the position of the Fermi level at the interface expected from direct band alignment. These important effects are revealed by our *ab initio* simulations of the full interface. For example, we find significant atomic disorder of Sb$_2$Te$_3$ near the interface with Cr, while Cr stays quite ordered (see Fig. 1a). (While the Seebeck coefficient in the disordered region may be changed compared to the rest of the TE material, the device Seebeck coefficient will not be strongly affected because the device Seebeck is an average over the thin film thickness, and the disordered region is only a very small fraction of the film thickness.)

The significant disorder at the interface is also accompanied by a strong chemical interaction between the Cr and the Sb$_2$Te$_3$. Fig. 1d shows the charge density difference between the Sb$_2$Te$_3$/Cr system and the isolated Sb$_2$Te$_3$ and Cr subsystems (having the same relaxed atomic positions as in the contact case). A large dipole with positive charge on the semiconductor forms within a few Angstroms from the metal interface. This dipole is opposite to the dipole formed due to the difference between the metal workfunction and the semiconductor electron affinity. The large values of the charge density difference reflect the strong interaction between metal and semiconductor atoms with chemical bonds at the interface. Farther away (more than 1 nm) from the metal interface the charge density difference acquires significant components from free carriers, which are responsible for the band bending as we shall see next. We note that the chemical interaction effect discussed here should be general and should be considered when studying the properties of other metal/semiconductor interfaces.

To further understand the influence of the metal on the electronic band structure of $Sb_2Te_3$, we calculated the spectral function projected on each of the six quintuplets included in the supercell (the term quintuplet refers to the group of five atomic planes that form one $Sb_2Te_3$ unit, see Fig. 1a.)[33]. The results are shown in Figure 2a. We find a completely distorted electronic structure in the first quintuplet, with states seen across the energy range and no discernible bandgap. The bulk semiconductor states become more clearly identifiable starting with the second quintuplet. In particular, we can identify the Fermi level positioned at the edge of the conduction band, indicating that even though the semiconductor is p-type, there is strong n-type doping near the interface due to the metal. One can see that the VBM and CBM start bending up as one moves further away from the interface, but with the VBM still below the Fermi level even in the 5$^{th}$ quintuplet. As we shall see, this is an artifact of the limited k-point sampling that we can employ in our full *ab initio* calculations. A similar picture holds for the undoped $Sb_2Te_3$ contacted with Cr, except that here the VBM and CBM are generally aligned about 50-100 meV lower in energy. We also note that the 6$^{th}$ quintuplet is next to vacuum in the simulations, hence the appearance of the topological surface states (Dirac cone) in the projected bandstructure.

We obtain the band bending $V(\vec{r})$ for the $Sb_2Te_3$/Cr system by subtracting the electrostatic potential of the semiconductor and metal subsystems from the one of the full metal/semiconductor system. We do not find an overall tunnel barrier at the interface as short bond-lengths between metal and semiconductor atoms allow for conduction paths near the Fermi level within a few Å from interface.

Fig. 2b shows the *ab initio* band-bending (planar-averaged along directions parallel to the layers/interface) for both doped and undoped $Sb_2Te_3$ contacted by Cr. The potentials represent the mid-gap of the semiconductor which bends up away from the interface due to free carrier accumulation in the valence/conduction band. In each case we have anchored the potential with respect to the Fermi level such that at d~5 nm the VBM corresponds to the value seen in the projected bandstructure (here we have assumed a bandgap of 130 meV as deduced from bulk calculations with the same k-point sampling).

Having obtained the detailed properties of the interface from ab initio calculations, we employ a macroscopic model that captures these details while allowing the modelling of band bending far away from interface together with the prediction of the contact resistivity (although fully *ab initio* approaches to obtain the contact resistivity have been developed[34] the size and complexity of this system prohibits the use of such approaches.) Since our novel multiscale approach is general, we expect it to be applicable to a broad range of metal/semiconductor interfaces. As discussed in the Supplementary Information, we self-consistently solve for the continuous charge distribution and electrostatic potential for spatially uniform p-type doping. We use a parabolic band approximation with effective mass obtained from the *ab initio* simulations. We also include the atomistic charges obtained from the *ab initio* simulations which arise from the strong chemical interaction at the interface. As shown in Fig. 2b the combined *ab*

*initio*/macroscopic approach can be used to predict the band bending in excellent agreement with the fully *ab initio* one in both the doped and undoped cases. Our approach allows to converge the band bending up to large distances away from the interface, as well as to consider arbitrary doping levels. The converged result for a doping of 6.5e19 holes/cm$^3$ is shown in Fig. 2c. Remarkably, we find that the band bending near the interface is mostly dominated by the chemical dipole charges, i.e. the p-type doping matters little near the interface and in fact the semiconductor is strongly n-type up to 3 nm away from the interface. Moving away from the interface, the chemical dipole charge decreases and quickly becomes smaller than the free carrier charge due to doping, leading to a band-bending determined by the doping (as can be deduced from the leveling of the electrostatic potential due to just the chemical dipoles $V_{dipole}$ as shown in Fig. 2c). The semiconductor becomes p-type at 5 nm from the interface. Thus, hole transport takes place through an effective n-p junction inside the semiconductor: holes injected from the contact are thermally excited over a Schottky barrier equal to the semiconductor bandgap (thermionic emission), or tunnel through a barrier of width W = 2 nm (thermionic field emission). This situation is illustrated in Fig. 3a.

We calculate the contact resistivity using conventional models for tunneling and thermionic emission[35] as detailed in the Supporting Information, with the band-bending for different doping obtained from the macroscopic model benchmarked with the *ab initio* simulations. We find that the barrier width is inversely proportional to the doping level (see Supporting Information), with negligible dependence on temperature. Other parameters are obtained directly from the *ab initio* simulations: the tunneling effective mass and barrier height. Figure 3b shows the calculated contact resistivity as a function of p-type doping for several temperatures. Three transport regimes can be identified: at high doping, the tunneling width is small and direct tunneling near the Fermi level dominates transport. At intermediate doping we find that the main contribution comes from thermionic field emission. Finally, at low doping hole transport takes place via thermionic emission, hence the stronger dependence of $\rho_C$ on temperature. At room temperature and for doping levels similar to those realized experimentally[18] (namely 3-5 x$10^{19}$ holes/cm$^3$) we find values of $\rho_C$ in the range 1- 3 x$10^{-8}$ $\Omega$cm$^2$, up to two orders of magnitude less than previously reported experimentally[18]. In addition to establishing a lower limit for the contact resistivity of realistic interfaces, our results also establish a benchmark to which future contact design modeling can be compared. Our results also provide the quantitative dependence of the contact resistivity on doping, which will have to be taken into account when optimizing the doping of future TE thin film devices.

Under the guidance of theoretical predictions we have tested experimentally the possibility of low $\rho_C$ by utilizing an improved technique to fabricate electrical contacts to V-telluride materials. The samples to be metallized were a p-type $Bi_2Te_3$/$Sb_2Te_3$ thin-film superlattice and an n-type $Bi_2Te_{3-x}Se_x$ thin-film alloy. Both type of films were grown epitaxially via metal-organic chemical vapor deposition on semi-insulating GaAs substrates[2]. These materials were loaded into a vacuum chamber and pumped down to a background pressure of less than 9x$10^{-7}$ torr. They were then back-sputtered using an Ar$^+$ plasma (power = 350W; time = 10 minutes; pressure = 20 mTorr; Ar flow rate = 50 sccm) in an effort to treat the sample surfaces in situ. Finally, the desired metal stacks were deposited onto the samples via sputtering with an Ar$^+$ plasma. Previous work has shown that similar processes resulted in decreased contact resistivity for Ni and Co contacts to $Bi_2Te_3$[36]. Schematics of the metal stacks can be found in the Supplementary Information. This surface treatment and metallization process can also be used to fabricate contacts onto other TE materials, such as PbTe and skutterudites, although the effectiveness of this technique on such materials is unknown.

The measured values of specific contact resistivity for the sputtered metallizations were obtained via the circular transfer length method (C-TLM)[37]. C-TLM calls for the measurement of electrical resistance values of a metallized thin film material across a series of annular gap spacings as a function of gap width (see Supplementary Information for details of the geometry). As the gap width approaches zero, the measured electrical resistance approaches the value of the contact resistance. Representative resistance vs. gap spacing data for both p-type and n-type V-telluride materials in given in Figure 4. The slope of the resistance vs. gap spacing line is the sheet resistance ($R_s$) multiplied by a geometric constant. The intercept is -2 times the transfer length ($L_T$), where the transfer length is the average physical distance an electron will travel in the semiconductor under the contact before flowing into the contact itself[38]. The specific contact resistivity is then calculated as $R_s * L_T^2$.

The specific contact resistivity values obtained for a number of sputtered metallizations, on both p-type and n-type V-telluride materials, are given in Table 1. Missing data in Table 1 means the calculated value resulted in a negative contact resistivity or transfer length, which is an unphysical result and therefore not reported. The traditional interpretation of this unphysical result is that the particular contact resistivity (and therefore transfer length) being measured is small compared to experimental variance, implying a very small value of the contact resistivity – likely below $10^{-8}$ ohm-cm$^2$. This data indicates that the Ar$^+$ backsputter/sputtered metallization process can yield lower $\rho_C$ values than the standard evaporation or plating metallization processes which are currently used in state-of-the-art thin-film V-telluride devices[18]. Indeed, we measured the properties of similar contact prepared without Ar sputtering and consistently find contact resistivities around $10^{-6}$ ohm-cm$^2$ for p-type contacts $10^{-7}$ ohm-cm$^2$ for n-type contacts, as shown in the Supporting Information.

To shed further light on the impact of Ar$^+$ backsputtering on the contact properties, we performed a scanning transmission electron microscopy (STEM) analysis of the Cr/p-type (Bi,Sb)$_2$Te$_3$ material. TEM specimens were prepared by focused ion-beam (FIB) liftout techniques and were analyzed using a Titan$^{TM}$ G2 80-200 instrument (FEI, Hillsboro OR, USA) operated at 200 keV and equipped with a SuperX$^{TM}$ window-less x-ray detector array for energy dispersive x-ray spectroscopic (EDS) analysis. EDS spectrum images were post-processed using multivariate statistical analysis methods using the Sandia-developed Automated eXpert Spectrum Image Analyzer (AXSIA) software[39].

Figure 5 shows STEM images and an EDS spectrum image of a sputtered Cr metal contact on a p-type (Bi,Sb)$_2$Te$_3$ superlattice[12]. The Cr layer here is nominally 40 nm thick and is topped with an overlayer of Au. The grains in the Cr layer exhibit a columnar morphology and, as can be seen in the corresponding EDS spectrum image, some Au appears to have infiltrated along the grain boundaries in this layer. However, while metal diffusion into the TE is a well-known phenomenon, here we find little evidence of metal diffusion into the TE material. Particularly striking is the rough morphology of the interface between the (Bi,Sb)$_2$Te$_3$ superlattice and the Cr coating. We believe that this rough morphology is likely a result of the pre-sputtering of the thermoelectric surface prior to deposition of the Cr. Analysis of the EDS spectrum images in conjunction with atomic resolution HAADF-STEM imaging (Figure 5(b)) shows vertical spikes of (Bi,Sb)$_2$Te$_3$ in contact with Cr, at least at their tips. We also identify an additional signal in the troughs (shaded as magenta in Fig. 5c) which exhibits signatures of enriched oxygen content (Fig. 5d). The observations also suggests some disorder in crystal structure of the (Bi,Sb)$_2$Te$_3$ material near the tip of the vertical spikes, with bending of the lattice planes and a disruption of the normal 5-plane wide quintuple layer structure apparent in the high resolution STEM observations (Fig. 5b). We also observe a similar morphology at the interface between the $(Bi,Sb)_2Te_3$ and the Ti metallization (see Supplementary Information). Overall, these observations suggest that controlling the interface morphology may be beneficial to the contact resistivity, and also point to possible further reduction of the contact resistivity via increasing the direct contact area by further removing the oxygen at the interface, which could allow the resistivity to approach the fundamental theoretical limit.

We harnessed the low contact resistivity of sputtered contacts to V-telluride materials by realizing high-performance thermoelectric coolers. A sputtered Ti/Cu metallization with an $Ar^+$ backsputter was used for the sink-side contact in a thin-film TE test structure, as shown schematically in Figure 6. The p-type material is a $Bi_2Te_3/Sb_2Te_3$ superlattice (thickness 15.3 μm) while the n-type material is $Bi_2Te_{3-x}Se_x$ (thickness 14.2 μm). Details of the material properties can be found in the Supplementary Information.

The hot side/cold side temperature difference ($\Delta T$) of the test device was measured as a function of input current. The top and bottom TE module temperatures, $T_C$ and $T_H$, respectively, were read using 25 μm diameter K-type thermocouples positioned on the device. The measurements were taken under vacuum ($P < 10^{-4}$ Torr) to minimize convective parasitics. The TE device was sunk to a water-cooled heat sink which was maintained at 23ºC. During the measurements, the temperature of the hot side increased slightly, reaching 24.5ºC at maximum cooling.

The $\Delta T$ vs. I curve for the Ti/Cu-contacted device is plotted in Figure 6 (additional data can be found in the Supplementary Information). Data from a control sample using standard contacts on both the source and sink sides is included for comparison. Both curves follow the expected expression[2]

$$\Delta T = \frac{ST_c}{K}I - \frac{1}{2K}(R + R_c)I^2 \quad (1)$$

where S is the Seebeck coefficient, $T_c$ is the cold side temperature, I is the current, K is the thermal conductance, R is the resistance of the thermoelectric material, and $R_c$ the contact resistance. (In our measurements, we stopped the data acquisition soon after ΔT started to decrease after reaching the maximum value. Best fits to Eq. (1) are shown to help guide interpretation of the curves.) Equation 1 shows that for fixed material parameters, a decrease in contact resistance leads to a decrease of the Joule heating term. This reduction allows the device to be operated at a higher current, and therefore to achieve a larger $\Delta T_{max}$.

The increased maximum ΔT value from the device built with sputtered Ti/Cu contacts ($\Delta T_{max}$ = 41.2K) as compared to that from the device with standard contacts ($\Delta T_{max}$ = 35.9K) suggests than the decreased $\rho_C$ seen from the sputtered Ti/Cu contacts can be readily translated into improved device performance. The reduction in contact resistance is readily seen from the voltage vs I data shown in the Supplementary Information, which gives total resistances of 18 mOhm for the reference module, and 10 mOhm for the module with sputtered contacts. Thus, adaptation of this new metallization process has the potential to enable record-setting performance of V-telluride devices, specifically those with thermoelectric applications. From the best fits and the above total resistances, we extract device Seebeck coefficients of 343 µV/K for the reference device, and 295 µV/K for the device with sputtered contacts. These values are in good agreement with the materials Seebeck coefficients reported in the Supplementary Information. In addition, this further shows the value of the improved contacts, since at the same device Seebeck coefficient the maximum cooling with the new contacts would be even higher.

In summary, we employed *ab initio* simulations to understand the fundamental properties of metal/telluride interfaces and to predict their contact resistivity. We find an unusual situation where the strong disorder at the metal/Te interface leads to chemical dipoles that dominate the band bending near the interface. This effect, accompanied by the high electron affinity of the TE material invariably leads to a contact resistivity that is determined mainly by thermionic field emission. We harness this knowledge to fabricate improved contacts to TE materials, and to demonstrate a significant improvement in the cooling performance of thin film TE devices. We expect that detailed knowledge of the contact properties could lead to further reduction of the contact resistivity by designing new contact geometries. For example, the knowledge that tunneling plays a critical role at the contact suggests that metal protrusions could provide local field concentration that would reduce the band bending width and increase the tunneling probability.

**Acknowledgements**

This work was supported by DARPA/MATRIX. Sandia National Laboratories is a multi-program laboratory managed and operated by Sandia Corporation, a wholly owned subsidiary of Lockheed Martin Corporation, for the U.S. Department of Energy's National Nuclear Security Administration under contract DE-AC0494AL85000.

REFERENCES

[1] B. Poudel, Q. Hao, Y. Ma, Y. Lan, A. Minnich, B. Yu, X. Yan, D. Wang, A. Muto, D. Vashaee, X. Chen, J. Liu, M. S. Dresselhaus, G. Chen, Z. Ren. *Science* **2008,** *320,* 634.

[2] P. Barletta, E. Vick, N. Baldasaro, D. Temple. *Proceedings of the Fifteenth IEEE Intersociety Conference on Thermal and Thermomechanical Phenomena in Electronic Systems (ITherm)* **2016**, 1477.

[3] G. J. Snyder, E. S. Toberer. *Nat. Mater.* **2008,** *7*, 105.

[4] G. E. Bulman, E. Siivola, R. Wiitala, R. Venkatasubramanian, M. Acree, N. Ritz. *J. Electron. Mater.* **2009,** *38*, 1510.

[5] T. M. Tritt. *Annual Review of Materials Research* **2011,** *41*, 433.

[6] D. Hsieh, D. Qian, L. Wray, Y. Xia, Y. S. Hor, R. J. Cava, M. Z. Hasan. *Nature* **2008,** *452*, 970.

[7] D.-X. Qu, Y. S. Hor, J. Xiong, R. J. Cava, N. P. Ong. *Science* **2010,** *329*, 821.

[8] Z. Alpichshev, J. G. Analytis, J. H. Chu, I. R. Fisher, Y. L. Chen, Z. X. Shen, A. Fang, A. Kapitulnik. *Phys. Rev. Lett.* **2010,** *104*, 016401.

[9] Y. L. Chen, J. G. Analytis, J.-H. Chu, Z. K. Liu, S.-K. Mo, X. L. Qi, H. J. Zhang, D. H. Lu, X. Dai, Z. Fang, S. C. Zhang, I. R. Fisher, Z. Hussain, Z.-X. Shen. *Science* **2009,** *325*, 178.

[10] I. Chowdhury, R. Prasher, K. Lofgreen, G. Chrysler, S. Narasimhan, R. Mahajan, D. Koester, R. Alley, R. Venkatasubramanian. *Nat. Nanotechnol.* **2009,** *4*, 235.

[11] M. Tan, Y. Deng, Y. Hao. *J. Phys. Chem. C* **2013,** *117*, 20415.

[12] R. Venkatasubramanian, E. Siivola, T. Colpitts, B. O'Quinn. *Nature* **2001,** *413*, 597.

[13] T. C. Harman, P. J. Taylor, D. L. Spears, M. P. Walsh. *J. Electron. Mater.* **2000,** *29*, L1.

[14] Y. R. Koh, K. Yazawa, A. Shakouri. *Int. J. Therm. Sci.* **2015,** *97*, 143.

[15] D. M. Rowe, *CRC Handbook of Thermoelectrics*. CRC Press, Boca Raton London New York Washington, D. C., **1995**.

[16] D. Ebling, K. Bartholomé, M. Bartel, M. Jägle. *J. Electron. Mater.* **2010,** *39*, 1376.

[17] C.-N. Liao, C.-H. Lee, W.-J. Chen. *Electrochem. Solid St.* **2007,** *10*, P23.

[18] G. Bulman, P. Barletta, J. Lewis, N. Baldasaro, M. Manno, A. Bar-Cohen, B. Yang. *Nat. Commun.* **2016,** *7*, 10302.

[19] D. Kraemer, J. Sui, K. McEnaney, H. Zhao, Q. Jie, Z. F. Ren, G. Chen. *Energy Environ. Sci.* **2015,** *8*, 1299.

[20] J. Maassen, C. Jeong, A. Baraskar, M. Rodwell, M. Lundstrom. *Appl. Phys. Lett.* **2013,** *102*.

[21] F. Kalaitzakis, G. Konstantinidis, L. Sygellou, S. Kennou, S. Ladas, N. Pelekanos. *Microelectron. Eng.* **2012,** *90*, 115.

[22] F. G. Kalaitzakis, N. T. Pelekanos, P. Prystawko, M. Leszczynski, G. Konstantinidis. *Appl. Phys. Lett.* **2007,** *91*, 261103.

[23] M. C. Shaughnessy, N. C. Bartelt, J. A. Zimmerman, J. D. Sugar. *J. Appl. Phys.* **2014,** *115*, 063705.

[24] J. Manson, A. Madubuonu, D. A. Crandles, C. Uher, P. Loš'ták. *Phys. Rev. B* **2014,** *90*, 205205.

[25] L. B. Duffy, A. I. Figueroa, Ł. Gładczuk, N. J. Steinke, K. Kummer, G. van der Laan, T. Hesjedal. *Phys. Rev. B* **2017,** *95*, 224422.

[26] T. Yamauchi, A. Kinoshita, Y. Tsuchiya, J. Koga, K. Kato In *1 nm NiSi/Si Junction Design based on First-Principles Calculation for Ultimately Low Contact Resistance*, 2006 International Electron Devices Meeting, 11-13 Dec. 2006, 2006; pp 1.

[27] T. Markussen, K. Stokbro In *Metal-InGaAs contact resistance calculations from first principles*, 2016 International Conference on Simulation of Semiconductor Processes and Devices (SISPAD), 6-8 Sept. 2016, 2016; pp 373.

[28] D. Vaidya, S. Lodha, S. Ganguly. *J. Appl. Phys.* **2017,** *121*, 145701.

[29] M. Aldegunde, S. P. Hepplestone, P. V. Sushko, K. Kalna. *Journal of Physics: Conference Series* **2015,** *647*, 012030.

[30] G. Kresse, J. Furthmüller. *Comput. Mater. Sci.* **1996,** *6*, 15.

[31] G. Kresse, J. Furthmuller. *Phys. Rev. B* **1996,** *54*, 11169.

[32] R. Kim, S. Datta, M. S. Lundstrom. *Journal of Applied Physics* **2009,** *105*, 034506.

[33] C. D. Spataru, F. Léonard. *Phys. Rev. B* **2014,** *90*, 085115.

[34] S. V. Faleev, F. Léonard, D. A. Stewart, M. van Schilfgaarde. *Phys. Rev. B* **2005,** *71*, 195422.

[35] R. Stratton. *J. Phys. Chem. Solids* **1962,** *23*, 1177.

[36] R. P. Gupta, K. Xiong, J. B. White, K. Cho, H. N. Alshareef, B. E. Gnade. *J. Electrochem. Soc.* **2010,** *157*.

[37] J. Klootwijk, C. Timmering. *Proceedings of the 2004 International Conference on Microelectronic Test Structures* **2004,** *17*, 247.

[38] D. Schroeder, *Semiconductor Material and Device Characterization*. Wiley-IEEE Press, **2015**.

[39] P. G. Kotula, M. R. Keenan. *Microsc. Microanal.* **2006,** *12*, 538.

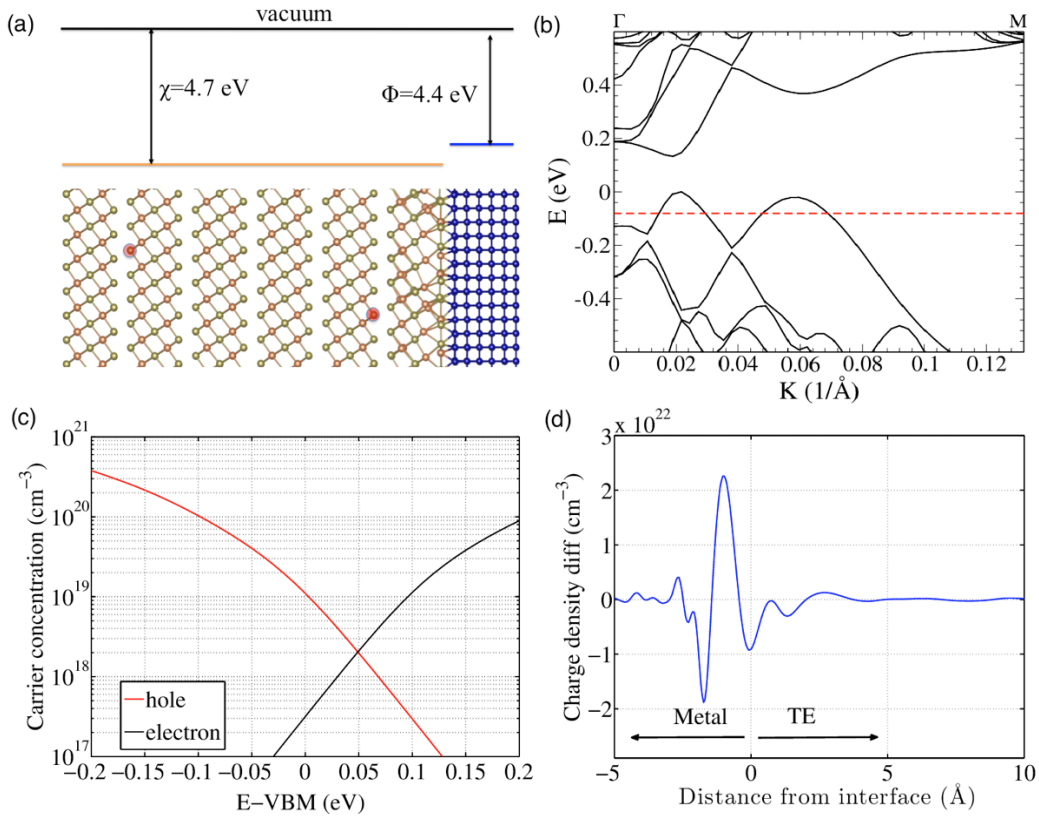

**Figure 1** Structure and electronic properties of interface. (a) Schematic diagram of band bending using cenventional band alignement. (b) Electronic band structure of $Sb_2Te_3$ for wave-vectors along the $\Gamma$ to M direction. Horizontal dashed line indicates the Fermi level for the p-type doping considered in the *ab initio* calculations. (c) Hole and electron carrier concentration as function of Fermi level in bulk $Sb_2Te_3$. (d) Charge density difference (mixed system minus metal and TE alone) at the $Cr/Sb_2Te_3$ interface.

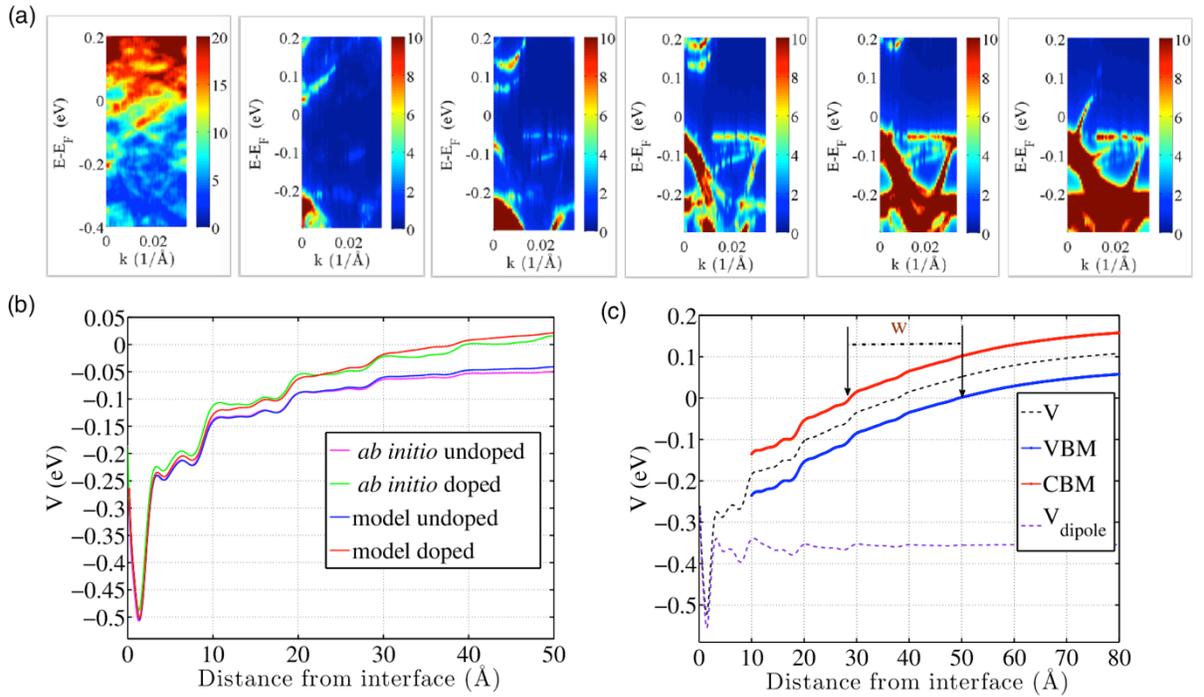

**Figure 2** Electronic structure and band-bending. (a) Projected spectral function on each quintuplet layer for p-doped interface of $Sb_2Te_3$ and Cr (the leftmost panel corresponds to the 1st quintuplet layer -near Cr, the rightmost to the 6th one –near vacuum). (b) Band bending in $Sb_2Te_3$ near the Cr interface calculated from first principles using a k-point sampling 2x1x1 compared with the ones obtained via macroscopic modeling using the corresponding *ab initio* charge carrier densities. (c) Band bending as well as VBM and CBM band profiles obtained via macroscopic modeling using the converged *ab initio* charge carrier densities shown in Fig. 1c. Also shown is the contribution to the band bending potential from chemical dipoles near the interface.

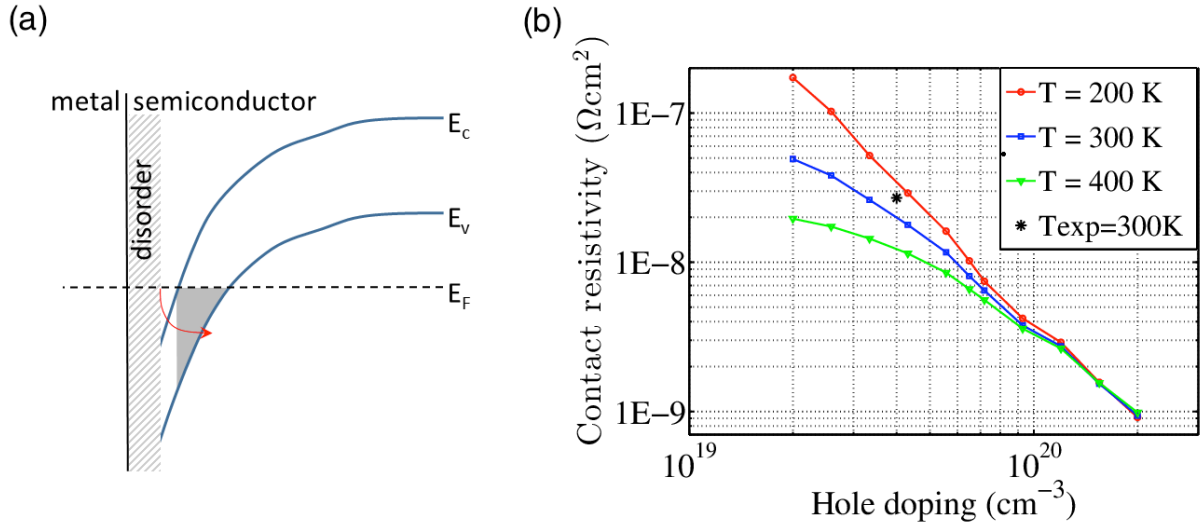

**Figure 3** Contact resistivity. (a) Illustration of the band bending leading to thermionic field emission from the metal to the semiconductor. The barrier width and height are obtained from macroscopic simulations benchmarked with *ab initio* simulations. (b) Calculated contact resistivity as a function of doping for three temperatures. The experimentally measured contact resistivity at room temperature is indicated with the symbol (the hole doping level was obtained via Hall measurements).

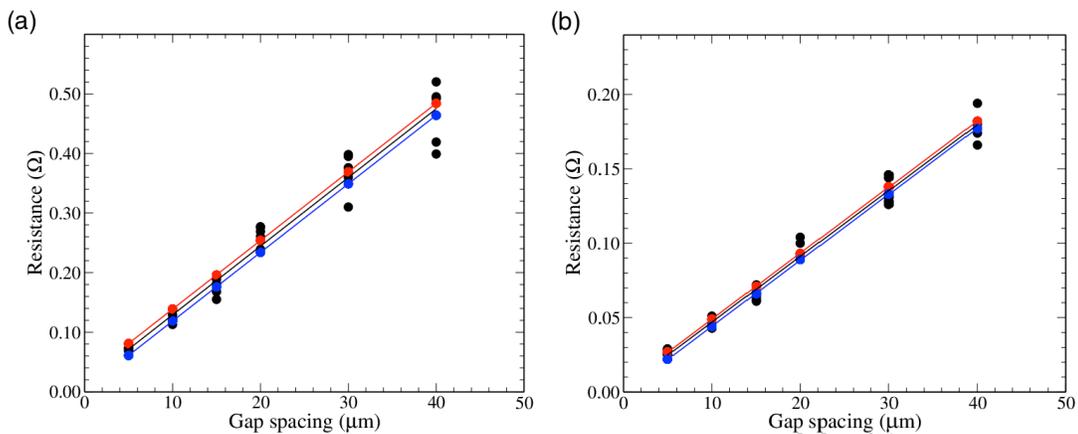

**Figure 4**. Resistance vs. gap spacing data for p-type (a) and n-type (b) V-telluride material that has been metallized with Ti via sputtering after an *in situ* Ar$^+$ ion backsputter treatment. The black

markers represent measured electrical resistance values. The red and blue markers represent +2σ and -2σ values, respectively. The lines represent least-squares fits to the data.

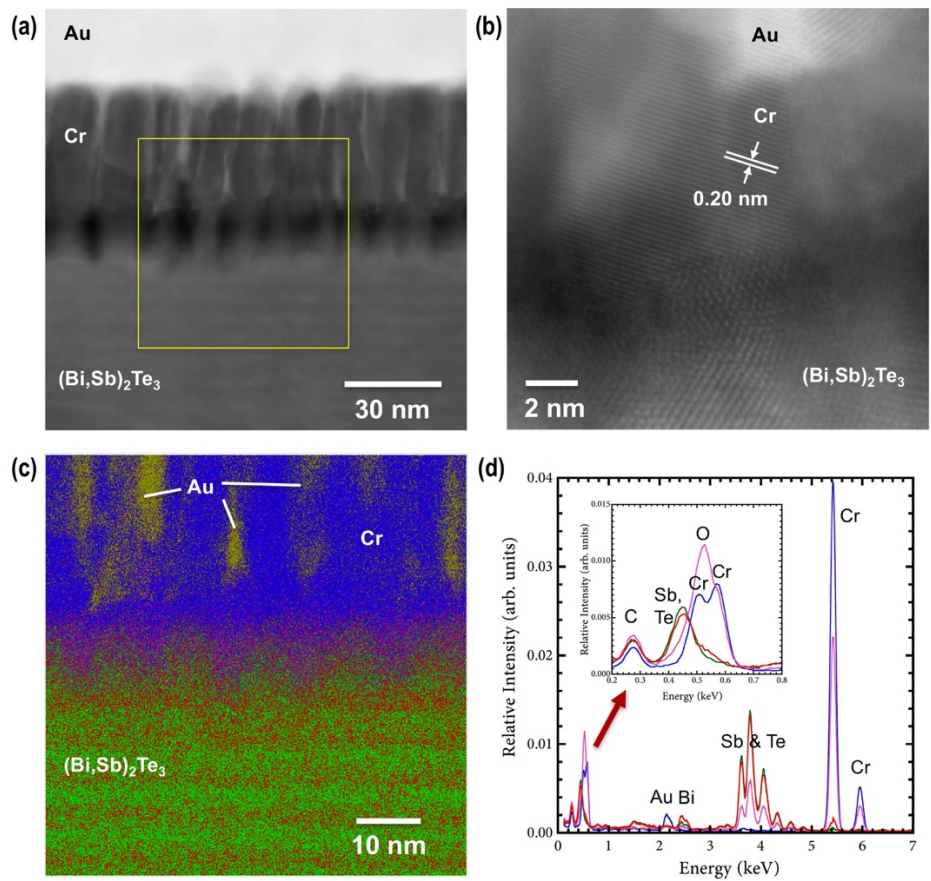

**Figure 5.** (a) High Angle Annular Dark Field (HAADF) STEM image showing microstructure of the sputtered Cr metallization on a $(Bi,Sb)_2Te_3$ superlattice. (b) HAADF STEM image showing atomic resolution of the interfacial region near the tip of one of the vertical $(Bi,Sb)_2Te_3$ spikes. Here, the Cr metal layer is in direct contact with the $(Bi,Sb)_2Te_3$. The measured 0.20 nm lattice fringe spacing in the metal layer is consistent with the spacing of the Cr (011)-type planes. (c) EDS spectrum image collected and analyzed using the multivariate curve resolution (MCR) technique from the region indicated by the yellow box in (a). 5 distinct spectral components were identified as follows: Red and Green correspond to Bi-rich and Bi-deficient layers in the $(Bi,Sb)_2Te_3$

superlattice; Blue corresponds to the Cr metallization. Yellow represents Au-rich regions within the Cr layer. Magenta is an oxygen-rich interlayer between the Cr and $(Bi,Sb)_2Te_3$. (c) Plot of the EDS spectral components using the same color scheme as in (c). For clarity, the component associated with Au is not shown. The inset shows the low energy region of the spectra. The interfacial phase (magenta) is associated with an increase in oxygen signal.

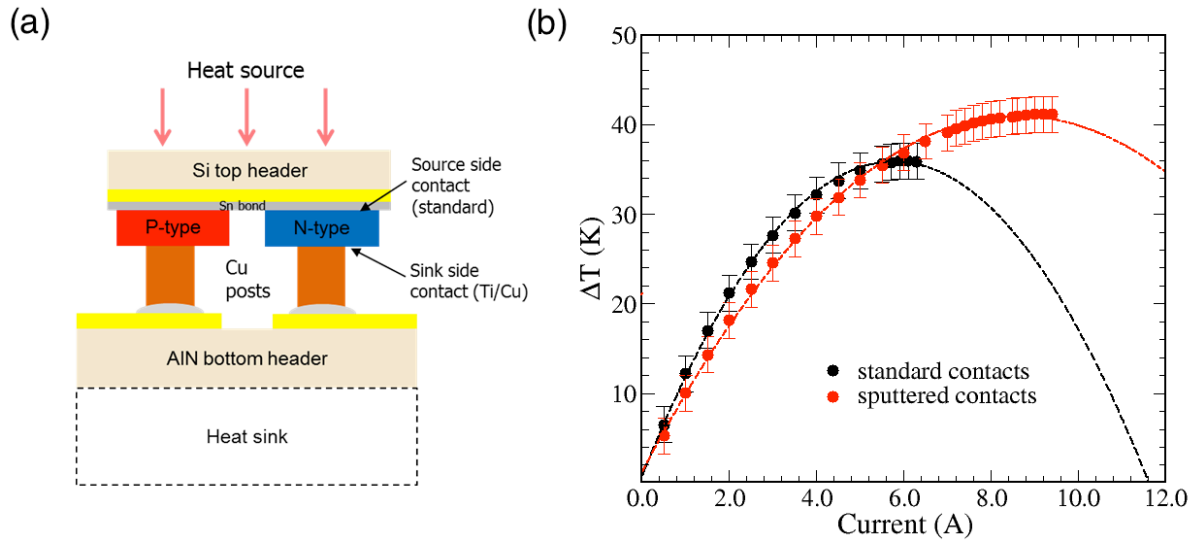

**Figure 6**. (a) A schematic of the test device used in this study. The sink-side contact was sputtered Ti/Cu with an $Ar^+$ backsputter. P-type and n-type active TE layers are 15μm in thickness. Figure is not drawn to scale. (b) ΔT vs. current plot for the thin-film thermoelectric coolers fabricated in this study. The red dots represent data from the device with sputtered Ti/Cu contacts. The black dots represent data from the device with standard contacts. The dashed lines are best fits to the data using Eq. (1).

**Table 1.** Transfer length and specific contact resistivity values for a series of metallizations for both p- and n-type V-telluride materials. An error of ±2σ has been applied to the data and is also given in the table.

| Type | Sputtered metallization | Transfer length, $L_T$ (μm) | | | Specific contact resistivity, $\rho_c$ ($\Omega cm^2$) | | |
|---|---|---|---|---|---|---|---|
| | | $+2\sigma L_T$ | $L_T$ | $-2\sigma L_T$ | $+\sigma_{\rho c}$ | $\rho_c$ | $-\sigma_{\rho c}$ |
| p | Au | 2.06 | 0.46 | -1.14 | 1.6E-07 | 2.6E-08 | -- |
| | Cr/Au | 0.84 | 0.53 | 0.21 | 4.9E-08 | 2.7E-08 | 4.4E-09 |
| | Cr/Ti/W/Au | 1.08 | 0.26 | -0.57 | 5.2E-08 | 9.3E-09 | -- |
| | Ti/Cu | 1.02 | 0.59 | 0.15 | 7.9E-08 | 3.9E-08 | -- |
| | Ti/TiW/Cu | 1.19 | 0.02 | -1.15 | 1.0E-08 | 1.3E-10 | -- |
| n | Au | -0.10 | -0.54 | -0.98 | -- | -- | -- |
| | Cr/Au | 0.16 | -0.09 | -0.34 | -- | -- | -- |
| | Cr/Ti/W/Au | 0.33 | 0.20 | 0.06 | 1.2E-08 | 6.1E-09 | 2.4E-10 |
| | Ti/Cu | 0.55 | 0.27 | -0.002 | 7.7E-09 | 3.2E-09 | -- |
| | Ti/TiW/Cu | -0.52 | -1.06 | -1.60 | -- | -- | -- |

# Supporting Information

**High Efficiency Thin Film Superlattice Thermoelectric Cooler Modules Enabled by Low Resistivity Contacts**

*Yuping He, François Léonard, Douglas L. Medlin, Nicholas Baldasaro, Dorota S. Temple[*],*
*Philip Barletta[*], and Catalin D. Spataru[*]*

**Table of Contents**

**S1. Details of *Ab Initio* simulations**

To understand the electronic transport properties at the interface between $Sb_2Te_3$ and Cr, we carried out a series of *ab initio* calculations. We generated the interfacial structure of $Sb_2Te_3$ and Cr by matching their experimental lattice constants. The primitive cell of $Sb_2Te_3$ has a rhombohedral structure with space group $D_{3d}^5$ ($R\bar{3}m$) and five atoms in one unit cell (i.e. three Te and two Sb). Its conventional unit cell is hexagonal with three ABC-stacked quintuplet layers and a total of fifteen atoms. On the other hand, Cr has the bcc crystal structure with space group $Im\bar{3}m$ and two atoms in one unit cell. In order to match the bcc structure of Cr, we converted the hexagonal unit cell of $Sb_2Te_3$ into a rectangular one, resulting in the lateral lattice constants $a_X$=4.26 Å and $a_Y$=7.38 Å. We created three interfacial structures consisting of $Sb_2Te_3$ contacted by the (100), (110) and (111) planes of Cr, and calculated the interfacial strain on Cr -we use the experimental lateral lattice constants for $Sb_2Te_3$- using different size of supercells. We found that the 4x4 $Sb_2Te_3$ supercell (we use 6 QPs along the Z-direction) and the 3x7 Cr (110) supercell (we use 6 atomic layers along the Z-direction) yield the smallest strain (i.e. 2% tensile strain in the X direction and 3% compressive strain in the Y direction). Because we found that all three Cr surfaces yield similar interface properties (disorder, band bending), we chose the one with the smallest strain for this study.

We then optimized the interfacial structure with smallest strain (4x4x2 $Sb_2Te_3$ and 3x7x3 Cr) with respect to the atomic position. The system has a total of 1464 atoms. The structure optimization was carried out using the VASP code[1], an *ab initio* simulation package based on Density Functional Theory. We used the Local Density Approximation (LDA) [2] and Projector Augmented Wave (PAW) [3]pseudo-potentials. To optimize the position of atoms we performed a Gamma-point only calculation. The relaxed structure has all the forces relaxed to less than 10 meV/Å. Previous studies [4] showed that the spin-orbit coupling (SOC) has a small effect on atomic structure, hence we did not include SOC in the structure optimization. When calculating the electronic structure of the optimized structure we sampled the Brillouin zone using a 2x1x1 k-point grid, we used a Fermi-Dirac smearing procedure with a temperature T=300K and we included SOC. To understand the effect of doping on the electronic transport at the interface, we also generated an interface of p-type doped $Sb_2Te_3$ and Cr by creating two Sb anti-site defects in the 4x4x2 supercell of $Sb_2Te_3$: one defect located between QL1 and QL2, the other one between QL5 and QL6. The positions of atoms around each defect within a radius of ~1 nm were further optimized.

**S2. Carrier concentration and density of transport modes from *ab initio* calculations**

We calculated the hole concentration as a function of Fermi level with respect to the VBM, using the formula $p = 1/V \sum_k (1 - f(E_k))$, where $f(E) = \frac{1}{1 + e^{(E-E_F)/k_B T}}$ is the Fermi Dirac distribution function, $E_F$ is the Fermi energy level, $k_B$ the Boltzmann constant, T the temperature, V is the volume of the supercell and $E_k$ are the electronic state energies of the valence band sampled with a large number of k-points (86x86x12 grid) over the whole Brillouin zone. We then calculated the density of transport modes (DOM) for hole carriers using the formula $M(E) = (h/2L) \sum_k |v_k^Z| \delta(E - E_k)$, where L is the length of supercell along the transport direction (Z), [5] and $\delta(E - E_k) = (1/w\sqrt{2\pi}) e^{-(E-E_k)^2/2w^2}$. The group velocities of the hole carriers were calculated by using finite difference method $v_k^Z = (1/h)(\Delta E_k / \Delta k_Z)$ with a small value of $\Delta k_Z$ along the Z direction. We used a small Gaussian smearing width w =5 meV. Similar expressions –with the sum being done over states from the conduction band - hold for the electron concentration and DOM.

Fig. S1 shows the DOM of bulk $Sb_2Te_3$ for a transport direction perpendicular to the Sb/Te layers, for both holes and electrons. The red dashed line represents a linear fit to the DOM for energies near the conduction band minimum (CBM). A transport electron effective mass $m_c$ can be obtained from the slope γ of the linear fit via [9]: $m_c = \gamma h^2 / 2\pi$ where h is the Planck constant. We obtain $m_c \approx 0.25\ m_0$, where $m_0$ is the bare electron mass. A similar value of hole effective mass $m_v$ is obtained at energies close to the valence band maximum (VBM). In our macroscopic modeling we only consider tunneling from CBM to VBM. Since $Sb_2Te_3$ has a quasi-direct band gap, electron-phonon scattering assistance is not required for momentum conservation parallel to the interface.

**S3. Converged band bending for large systems**

We obtain the band bending $V$ for the $Sb_2Te_3$/Cr system by subtracting the electrostatic potential of the semiconductor and metal subsystems from the one for the full metal/semiconductor system. This definition for $V$ results in a smooth function as the ionic potential contribution cancels out exactly. However, this definition also neglects the contribution of disorder and metallic surface states to the tunnel barrier and band bending. We have checked (via a macroscopic average [6] of the difference between the electrostatic potential of disordered semiconductor slab and ordered bulk semiconductor) that this contribution is negligible at distances larger than 1 nm away from the metal interface , which is the region of interest as the bulk states cannot be identified at shorter distances. The band bending is sensitive to the free carrier concentration as a function of the position of the Fermi level, and converges slowly with k-point sampling. Due to the large metal/semiconductor system (i.e. thousands of atoms) used in the *ab initio* calculations we could not use a k-grid denser than 2x1x1. The resulting density of states (DOS) with k=2x1x1 is such that the Fermi level is positioned near the VBM in the p-doped $Sb_2Te_3$, instead of being 0.75 meV below the VBM as we obtained from the calculated carrier concentration as a function of Fermi Level with a dense k-grid (86x86x12) for bulk $Sb_2Te_3$.

To address this issue and to predict the converged band bending for large systems, we employ a combination of macroscopic model and *ab initio* calculations. This is done in several steps:

i) we start from the *undoped ab initio* case noting that the charge density difference ρ and the band bending V are related to each other through the Poisson equation: $\nabla^2 V(z) = -\rho(z)$. The potential allows us to obtain the free carrier concentration ( i.e $n - p + N_d$ ) as a function of distance x away from the interface based on the fully *ab initio* results (but unconverged w.r.t. k-point sampling).

ii) we re-write the Poisson equation under the assumption that the response of the semiconductor valence bands to the perturbations due to free carriers can be integrated out by screening of the latter with the average macroscopic dielectric constant of the semiconductor ε (we use ε~80 based on *ab initio* results): $\nabla^2 V(z) = -\rho_0(z) + e[n(z) - p(z) + N_d]/\varepsilon$ where $\rho_0$ represents the charge localized near the interface due to chemical dipoles (which converges fast w.r.t. k-point sampling), $N_d$ is dopant density, n(z) and p(z) are electron and hole concentration, respectively. Here we also treat the Cr system as a perfect metal. Comparison with step i) allows to obtain $\rho_0(z)$ from which one can predict the band bending for other cases.

iii) we benchmark our modeling by calculating the band bending for the *doped* case using *ab initio* carrier concentrations obtained from 2x1x1 k-point sampling. The resulting potential is shown in Fig. 2b. We see that the modeling reproduces very well the *ab initio* calculated band bending at distances larger than ~2nm from the interface. At shorter distances a slight difference exists which we believe is due to the fact that the electronic structure of bulk semiconductor (assumed in macroscopic modeling) is slightly different than the one of the disordered semiconductor slab (implied in the *ab initio* calculations) as well as the fact we idealize the metal contact.

iv) we use the combined *ab initio*/macroscopic modeling to predict the converged band bending in the doped case by re-calculating the solution of the Poisson equation in conjunction with the converged carrier concentration of bulk $Sb_2Te_3$ (See Fig. 1b and 1c) and interfacial charge due to chemical dipoles. The resulting band bending is shown in Fig. 2d, together with the band profile associated with VBM and CBM (assuming a band gap consistent with the converged k-grid, namely $E_g$=100 meV).

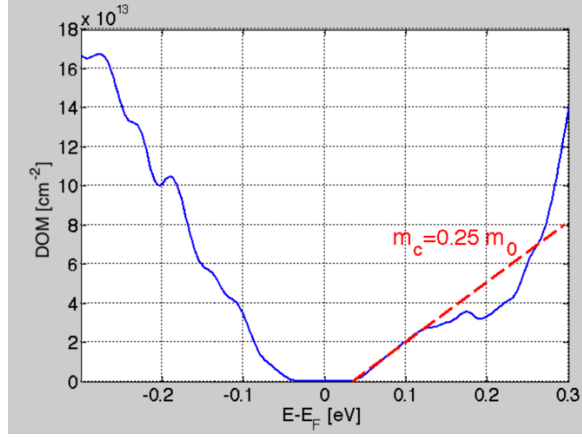

**Figure S1:** Calculated density of modes of bulk $Sb_2Te_3$ for transport direction perpendicular to Sb/Te layers.

## S4. Macroscopic modeling

We self-consistently solve Poisson's equation $\nabla^2 V(z) = -\rho_0(z) + e[n(z) - p(z) + N_d]/\varepsilon$ and the equations for the free carrier density

$$n(z) = \int_{E_c(z)}^{\infty} D_n(E,z) f(E - E_F) dE \quad (1)$$

$$p(z) = \int_{-\infty}^{Ev(z)} D_p(E,z) [1 - f(E - E_F)] dE \quad (2)$$

Here $\rho_0(z)$ is the (fixed) charge due to chemical dipoles obtained from the *ab initio* calculations, $N_d$ is the dopant concentration, and $D_n(E,z)$ and $D_p(E,z)$ are the electron and hole density of states locally shifted by the local electrostatic potential. This approach is the conventional rigid shift approximation.

The density of states follows an effective mass approximation

$$D_n(E,z) = \frac{8\pi\sqrt{2}m_e^{3/2}}{h^3}\sqrt{E - E_c(z)} \qquad (3)$$

$$D_p(E,z) = \frac{8\pi\sqrt{2}m_h^{3/2}}{h^3}\sqrt{E - E_v(z)} \qquad (4)$$

with the effective masses obtained from the *ab initio* calculations.

The equations are discretized on a one-dimensional dense grid and solved self-consistently using linear mixing at each iteration. Poisson's equation satisfies boundary conditions at the two ends of the simulation cell. At the metal/semiconductor interface we set a boundary condition V(z=0)=constant. Because the solution of Poisson's equation gives the potential up to a constant, we set the value of V(z=0) by matching macroscopic band-bending with the results of the *ab initio* calculations in Step 1, section S3. This is done by matching the VBM at the fifth quintuplet away from the interface, yielding a value V(z=0)=-0.27 eV. We note that V(0)~$\phi$-($\chi$+$E_g$/2) where $\phi$ is the metal work function and $\chi$ the semiconductor electron affinity (see Fig. 1a). Far away from the contact we impose charge neutrality $n(\infty) - p(\infty) + N_d = 0$ ; alternatively we can impose $\partial n(z)/\partial z = \partial p(z)/\partial z$ at z= $\infty$ . Both approaches give similar results. The Poisson equation can be solved numerically for any given doping with the above boundary conditions. The VBM and CBM level positions as function of distance from interface follow the electrostatic potential and we obtain the barrier width as $W = z_c - z_v$ where $E_c(z_c) = E_v(z_v) = E_F$. We find that W is inversely proportional to the doping level (see Fig. S2) with negligible dependence on temperature.

The contact resistivity is calculated from

$$\rho_c = \left(\frac{\partial J}{\partial V}\bigg|_{V=0}\right)^{-1} \quad (5)$$

with the zero bias current density given by[7]

$$\frac{\partial J}{\partial V}\bigg|_{V=0} = -\frac{4\pi me}{h^3}\int_0^\infty dE\, \frac{\partial f(E-E_F)}{\partial E}\int_0^E P(E_z)dE_z \quad (6)$$

and

$$P(E_z) = \exp\left[-\frac{2\sqrt{2m^*}}{h}\int \sqrt{eV(z)-E_z}\,dz\right] \quad (7)$$

This last equation is the tunneling transmission probability as a function of the component of the carrier energy $E_z$ perpendicular to the interface. Thermionic emission is included in our calulcations by setting $P(E_z)=1$.

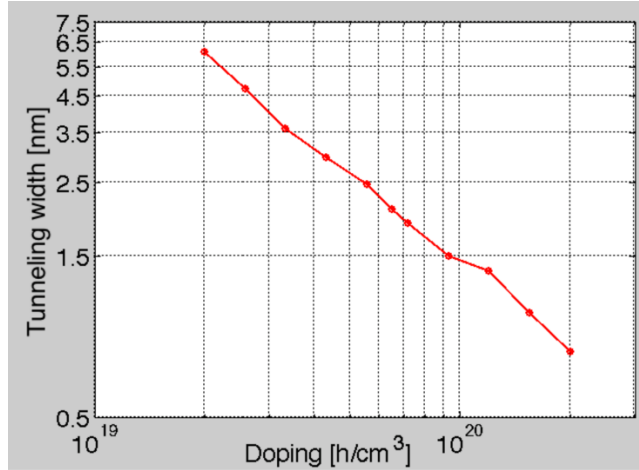

**Figure S2:** Tunneling width for hole transport across the p-n junction formed near the $Sb_2Te_3$-Cr interface as function of doping (T=300 K).

### S5. Thermoelectric properties of the $Bi_2Te_3$/$Sb_2Te_3$ superlattice material

Both of the modules were fabricated from the same p-type and n-type materials, which had thicknesses of 15.3μm and 14.2 μm, respectively. Unfortunately, there is no straightforward method to measure *cross-plane* properties for these materials in as-grown, on-substrate form. Therefore, we present the *in-plane* measurements for the Seebeck coefficient and electrical conductivity. This allows us to compare as-grown samples to each other. However, it is important to recognize that in an operational device, the material's *cross-plane* properties are what is relevant. Also, we did not measure the thermal conductivity for these two samples, but in the past, both of these materials have consistently given 1.0-1.1 W/m-K[8].

**Table S1:** Thermoelectric properties of the p-type and n-type materials used to measure contact resistance and to fabricate modules.

| p-type ($Bi_2Te_3/Sb_2Te_3$) | | | n-type ($\delta\text{-}Bi_2Te_{3-x}Se_x$) | | |
|---|---|---|---|---|---|
| Parameter | Value | Measurement Direction* | Parameter | Value | Measurement Direction* |
| Thickness | 15.3µm | N/A | Thickness | 14.2µm | N/A |
| Seebeck coefficient | 290 µV/K | In-plane | Seebeck coefficient | -355 µV/K | In-plane |
| Electrical resistivity | 7.2e-4 Ω-cm | In-plane | Electrical resistivity | 1.3e-3 Ω-cm | In-plane |
| Thermal conductivity | Not measured; 1.1 W/m-K estimated | Cross-plane | Thermal conductivity | Not measured; 1.1 W/m-K estimated | Cross-plane |

## S6. Voltage response of modules

The voltage response of the two modules is shown in Fig. S3. The lower overall resistance of the module with the lower contact resistance (blue data points) is clearly apparent from the comparison with the control module (red data points).

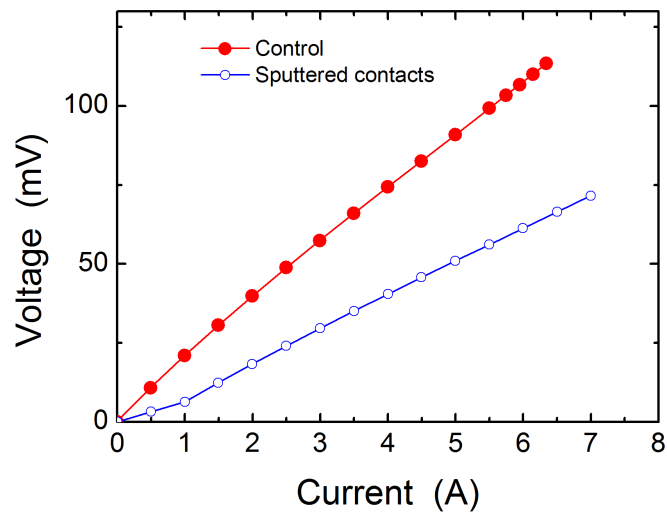

**Figure S3**: Voltage response of the two modules. The red data points are for the control module, the blue data points are for the module with improved contacts.

## S7. Contact resistivity of control samples

We measured the contact resistivity of control samples (i.e without Ar sputtering) using the same TLM technique described in the main text. The samples consisted of the same p-type and n-type thermoelectric materials as for the Ar sputtered contacts. Results are presented in Table S2.

**Table S2**: Contact resistivity of control samples.

| Leg | Side | Metallization | $\rho_c$ ($\Omega$-cm$^2$) |
|---|---|---|---|
| p-type | source | Evaporated Cr/Ni/Au | 1.42E-06 |
| | sink | Plated Au | 1.36E-06 |
| n-type | source | Plated Au | 2.68E-07 |
| | sink | Plated Au | 2.68E-07 |

## S8. Details of metal stacks and measurement geometry

Figure S4 shows the stacking details and dimensions for the different contact materials. We also include a schematic of the measurement geometry used to extrac the contact resistance[8].

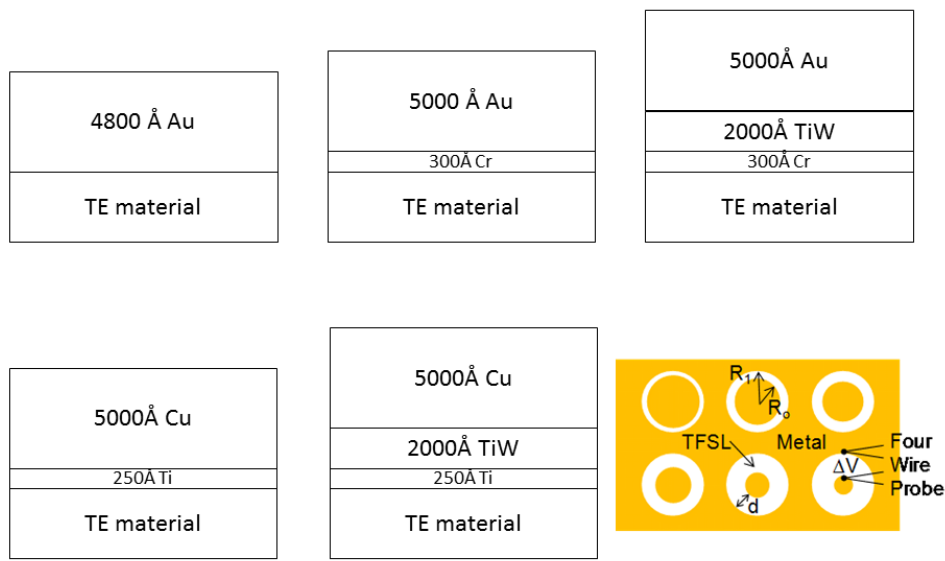

**Figure S4**: Cross-sectional schematics of the metal stacks used in the experiments. The bottom right panel shows the geometry used to obtain the contact resistance.

## S9. Microstructure of the Ti/(Bi,Sb)$_2$Te$_3$ interface

We also conducted STEM analysis of the interface between the Cu/Ti and p-type (Bi,Sb)$_2$Te$_3$ material (Figure S5). As with the Cr/(Bi,Sb)$_2$Te$_3$ we observe a rough interface, which we believe likely arises due to the Ar+ sputtering of the TE material. Also similar to the Cr/(Bi,Sb)$_2$Te$_3$ metallization, we observe a thin layer of oxygen enriched material at the interface.

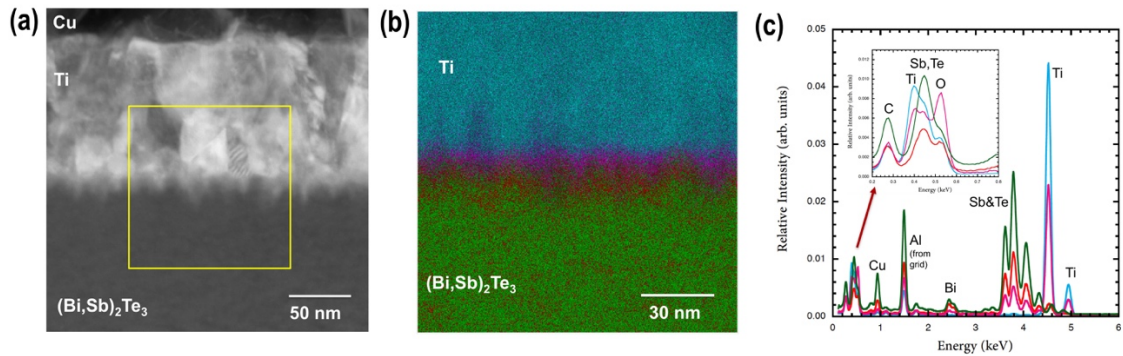

**Figure S5.** (a) Bright Field (BF) STEM image showing microstructure of sputtered Ti metallization on (Bi,Sb)$_2$Te$_3$ material. (b) EDS spectrum image from the region indicated by the yellow box in (a). Analysis was conducted using the multivariate curve resolution (MCR) technique as discussed in the main text. The spectral components, which are plotted in (c), are mapped with the following color scheme: Red and Green correspond to Bi-rich and Bi-deficient regions in the (Bi,Sb)$_2$Te$_3$ material; Teal corresponds to the Ti metallization; Magenta corresponds to an oxygen-rich interlayer at the interface between the Ti and (Bi,Sb)$_2$Te$_3$ material (Note: This specimen was mounted on an aluminum support grid, giving rise to additional Al signal in the spectra).

# References

[1]  G. Kresse, J. Furthmuller. *Phys. Rev. B* **1996,** *54*, 11169.

[2]  J. P. Perdew, A. Zunger. *Phys. Rev. B* **1981,** *23*, 5048.

[3]  P. E. Blochl. *Phys. Rev. B* **1994,** *50*, 17953.

[4]  H. Zhang, C.-X. Liu, X.-L. Qi, X. Dai, Z. Fang, S.-C. Zhang. *Nat Phys* **2009,** *5*, 438.

[5]  R. Kim, S. Datta, M. S. Lundstrom. *J. Appl. Phys.* **2009,** *105*, 034506.

[6]  R. G. Dandrea, C. B. Duke. *J. Vac. Sci. Technol. A* **1993,** *11*, 848.

[7]  R. Stratton. *J. Phys. Chem. Solids* **1962,** *23*, 1177.

[8]  G. Bulman, P. Barletta, J. Lewis, N. Baldasaro, M. Manno, A. Bar-Cohen, B. Yang. *Nat. Commun.* **2016,** *7*, 10302.